\documentclass [conference]{IEEEtran}

\usepackage{siunitx}
\usepackage{tabularx}
\usepackage{verbatim}
\PassOptionsToPackage{bookmarks={false}}{hyperref}
\usepackage{algorithm}
\usepackage{amsmath}
\usepackage{amsthm}
\usepackage{amssymb}
\usepackage{citesort}
\usepackage{fancyvrb}
\usepackage{algorithmic}
\usepackage[T1]{fontenc}
\usepackage[scaled]{helvet}
\usepackage{lscape}
\usepackage[dvips,final]{graphicx}
\usepackage{subfigure}
\graphicspath{{./}{./figures/}}
\usepackage{enumerate}
\usepackage{epstopdf}
\usepackage{float}
\usepackage{url}
\usepackage{caption}
\usepackage{subfig}

\IEEEoverridecommandlockouts

\begin{document}


\title{CAFT: Congestion-Aware Fault-Tolerant Load Balancing for Three-Tier Clos Data Centers \thanks{An IEEE-formatted version of this article is published in the  2020 International Wireless Communications and Mobile Computing (IWCMC) conference proceedings. Personal use of this material is permitted. Permission from IEEE must be obtained for all other uses, in any current or future media, including reprinting/republishing this material for advertising or promotional purposes, creating new collective works, for resale or redistribution to servers or lists, or reuse of any copyrighted component of this work in other works.}}

\author{Sultan Alanazi and Bechir Hamdaoui \\
\small Oregon State University, \small Corvallis \\
\small \{alanazsu,hamdaoui\}@eecs.oregonstate.edu~\\
}


\twocolumn
\maketitle 

\begin{abstract}
Production data centers operate under various workload sizes ranging from latency-sensitive mice flows to long-lived elephant flows. However, the predominant load balancing scheme in data center networks, equal-cost multi-path (ECMP), is agnostic to path conditions and performs poorly in asymmetric topologies, resulting in low throughput and high latencies.
In this paper, we propose CAFT, a distributed congestion-aware fault-tolerant load balancing protocol for 3-tier data center networks. It first collects, in real time, the complete congestion information of two subsets from the set of all possible paths between any two hosts. Then, the best path congestion information from each subset is carried across the switches, during the Transport Control Protocol (TCP) connection process, to make path selection decision. Having two candidate paths improve the robustness of CAFT to asymmetries caused by link failures. Large-scale ns-3 simulations show that CAFT outperforms Expeditus in mean flow completion time (FCT) and network throughput for both symmetric and asymmetric scenarios.

\end{abstract}

\begin{keywords}
Load balancing, data center networks, network congestion, distributed routing.
\end{keywords}

\section{\sc {Introduction}}
\label{sec:Introducation}
Today, modern cloud data center networks must be scaled to accommodate the increase in the number of tenants and applications. Recent papers such as VL2 \cite{vl2} and Fat-tree \cite{fat} present multi-rooted Clos topologies, enabling multiple paths between host pairs to provide high bisection bandwidth. Current data center operators load balance traffic using the equal-cost multi-path (ECMP) routing protocol~\cite{ecmp}. 
ECMP uses static hash functions to forward packets among equal cost outports. However, in ECMP, hash collisions cause applications to experience throughput degradation for throughput-sensitive flows and tail latency for delay-sensitive flows. 

Several network load balancing algorithms have been proposed to address the shortcoming of ECMP in literature. They can be broadly classified as centralized scheduling (e.g., Hedera \cite{hedera}), host-based load balancing (e.g., MPTCP \cite{mptcp}), and in-network congestion-aware load balancing (e.g., CONGA~\cite{conga} and Expeditus~\cite{expeditus}). Despite significant efforts, prior proposals still have important shortcomings. Centralized scheduling schemes are too slow to react to latency-sensitive mice flows. Host-based load balancing approaches such as MPTCP are not easy to deploy as the network operators do not have control over the end-host networking stack. Furthermore, MPTCP performs poorly in incast scenarios~\cite{conga}.

 In-network load balancing algorithms such as CONGA and Expeditus are distributed congestion-aware protocols. In CONGA, each Top-of-Rack (ToR) switch maintains end-to-end congestion information of all possible paths to other switches in the network. Congestion metrics are carried by data packet piggy-backing. CONGA is robust to asymmetry caused by link failure and performs well in 2-tier leaf-spine topology. However, it faces major scalability challenges in 3-tier Clos topologies. Expeditus, on the other hand, has been proposed to overcome CONGA's challenges, in 3-tier Clos networks, by presenting a two-stage path selection heuristic that uses partial path information for flow to path allocation. Nevertheless, Expeditus makes sub-optimal decisions in case of topological asymmetry because it only utilizes partial information of the possible paths during the decision making process.
 
 In this paper, we propose CAFT, an in-network congestion-aware fault-tolerant load balancing protocol for 3-tier Clos data center networks. CAFT overcomes the limitations of Expeditus design by collecting a complete link load information of two candidate paths before making load balancing decisions. It is a flow-based routing algorithm that allocates flows to paths during TCP's 3-way handshaking process without causing packet reordering. To summarize, our main contributions are as follows. We:
 \begin{enumerate}
  \item Propose a congestion-aware routing protocol that is robust to asymmetries caused by link failures.
  
  \item Evaluate our technique in large-scale networks using ns-3 simulator, and show that it performs better than existing schemes and close to the optimal.

\end{enumerate}
 
 The remainder is organized as follows. Section II explains our proposed protocol, CAFT. Section III evaluates our framework on large-scale network simulator. Finally Section IV concludes and provides directions for future work.

\section{\sc {CAFT}}
\label{sec:Framework}
\begin{figure}
	\centering{
\includegraphics[width=1\columnwidth]{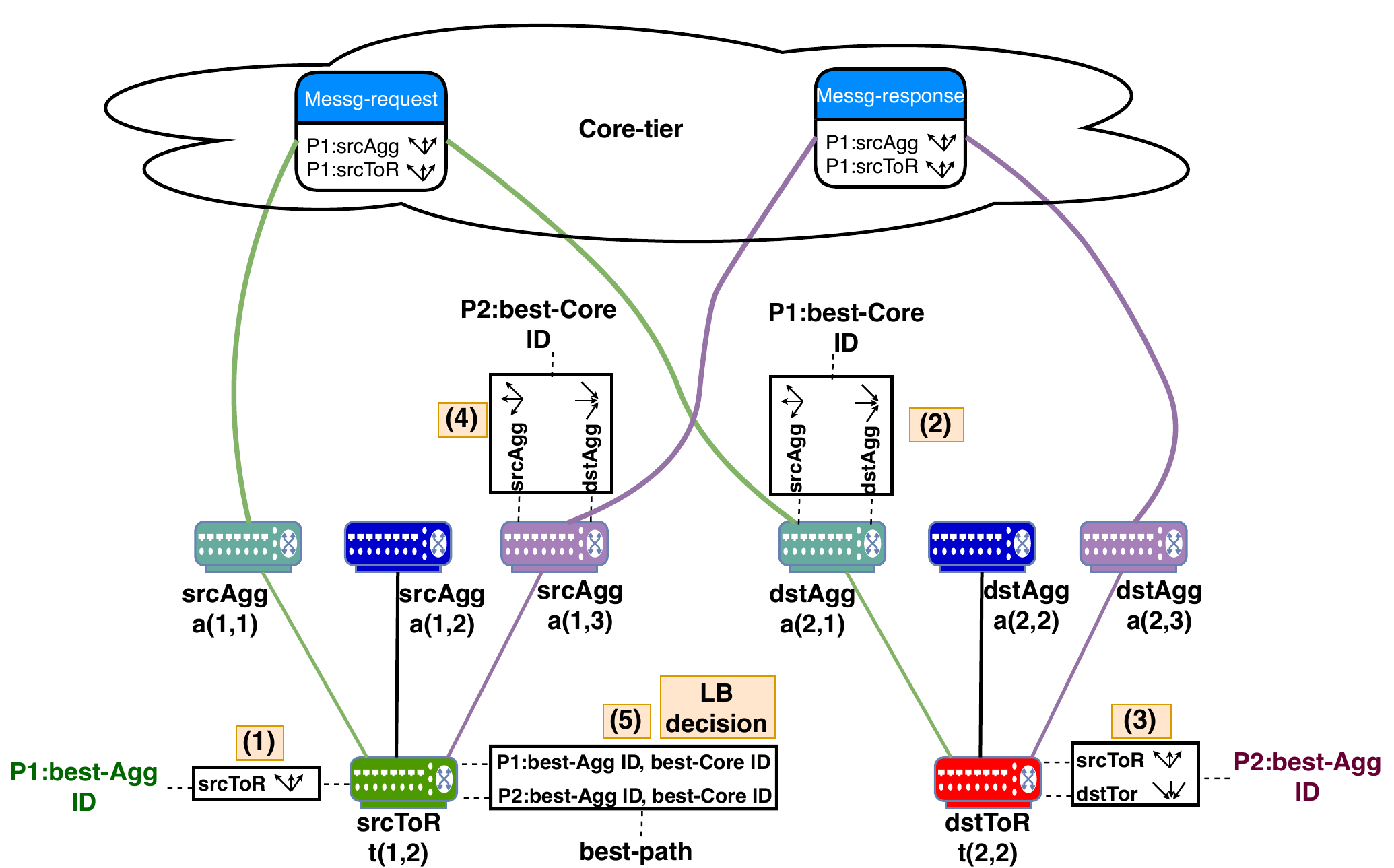}
	\caption{CAFT design. Four distributed decisions initiate two candidate paths during the TCP connection before making the final load balancing decision. }
\label{fig:Framework}}
\end{figure}

In this section we present CAFT, a distributed  \underline{C}ongestion-\underline{A}ware \underline{F}ault-\underline{T}olerant load balancing protocol for 3-tier data center networks. CAFT protocol is distributed in switches data plane. It is a flow-based routing mechanism that assigns flows to paths using link load as a metric for congestion. The work in \cite{conga} shows the efficiency of the congestion metric by implementing it in real hardware.
CAFT explores two candidate paths during the TCP handshaking process before making load balancing decision. Two switches are involved in the decision process of each path selection, ToR switch decides on the best path to the aggregation switch. As illustrated in figure~\ref{fig:Framework}, the green path decision is made by ToR $t(1,2)$ and aggregation $a(2,1)$ switches in steps $(1)$ and $(2)$; while for the purple path in steps $(3)$ and $(4)$, switches $t(2,2)$ and $a(1,3)$ made the path selection decision. In step $(5)$, the link load information of both paths are carried to the source ToR $t(1,1)$ which assigns a flow to the path that best balances the network load in the cluster.
To be congestion-aware and fault-tolerant, CAFT relies on two main components: information collection, where switches monitor links' congestion and failure, and round-trip based path selection, where the monitored information is aggregated across the network in a round-trip fashion to obtain two candidate paths for a flow to path allocation. Next, we provide in details the description of CAFT system's architecture:

\subsection{Key Components}
For a path to be selected in CAFT, ToR and aggregation switches are required to monitor their uplink ports that connect them to the higher level switches and record both the egress and ingress link load congestion metrics and the status of each link in case of failure. The recorded information will be readily available to be transported to other ToR and aggregation switches that are involved in the path selection process.

\subsubsection{Link Load Measurement}
The load of the link is estimated using Discounting Rate Estimator (DRE) \cite{conga}. The DRE keeps a register $X_{i}$ for each uplink port $i$, which is incremented by the packet size in bytes for each packet transmitted/received over the link, and is decremented every period $(T_{dre})$ with a factor $\alpha$ that ranges between $[0,1]$. The register X is proportional to the rate of the traffic over the link. The DRE is almost identical to the Exponential Weighted Moving Average (EWMA) which is widely used in practice. The parameter choices are set based on \cite{conga} where $T_{dre} = \SI{20}{\micro\second}$ and $\alpha = 0.1$. The congestion metric is quantized to 3 bits relative to the link speed.

\subsubsection{Congestion Table}
Switches maintain link congestion metrics in a congestion table using 4 bits where the first bit indicates a link failure while the remaining 3 bits represent the link utilization.

\subsubsection{Path Allocation Table (PAT)}
Flows to paths assignments are maintained in a PAT in each source ToR and aggregation switches, i.e the switches that connect the source host to the core switches. Once a core switch is selected to be part of a flow's path, there will be a unique path from that core to the destination host (structural property of 3-tie Clos). Each table entry represents a flow to path allocation and consists of a flow ID, the outport number, a valid bit indicating whether a flow is active (1) or not active (0), and an age bit indicating entry timeout. Upon the arrival of each packet, the switch looks up an entry based on a hash of its five-tuple (flow ID). If the entry exists and its valid bit equals 1, the packet is sent through the port indicated in the entry. If the entry is not valid or no entry exists, the incoming packet is either considered a new flow and triggers a new round-trip path allocation process (in case of source ToR switches), or forwarded using ECMP (in case of source aggregation switches). PAT entries time out if the flows are inactive for a long period of time indicating expired entries. Once an entry is expired, its valid bit is set to zero, triggering a new path allocation process. We follow \cite{conga} and set the time out value to 100ms as it is shown to be able to filter out bursts of the same flow. Only one bit is required to implement the timer for each entry. The work in \cite{conga} reported that the cost of tracking 64K flows is low for large scale data center networks. They have also estimated that a heavy loaded switch can have up to 8K concurrent flows.
\begin{figure*}[t]
	\centering
\includegraphics[width=1.95\columnwidth, height=6cm]{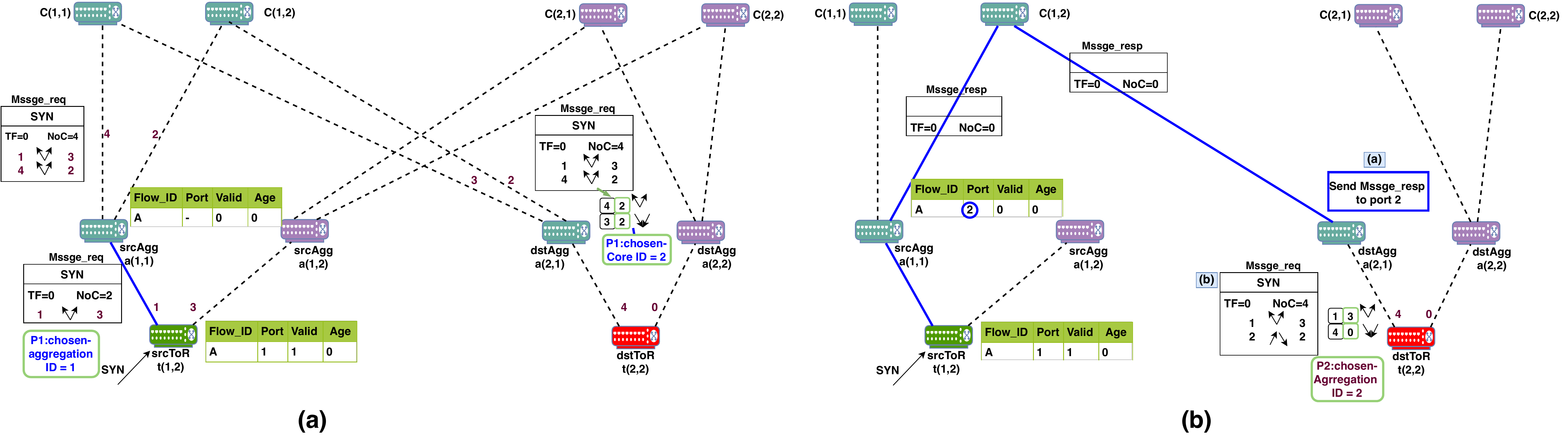}  
	\caption{First path selection in CAFT during the TCP handshaking process(a). The SYN packet is received by source ToR $t(1,2)$ and routed to destination ToR $t(2,2)$. Fig(b) illustrates the processing actions executed by destination aggregation $a(2,1)$. }
\label{fig:E1}
\end{figure*}

\begin{figure*}[t]
	\centering
\includegraphics[width=1.95\columnwidth, height=6cm]{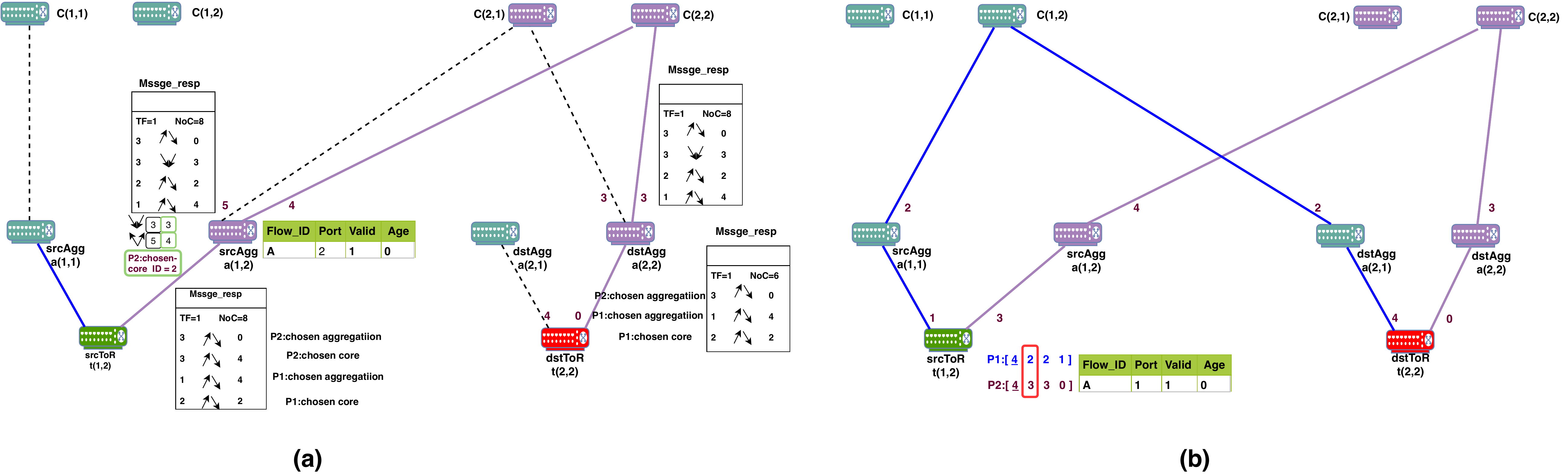}  
	\caption{Second path selection in CAFT during the TCP handshaking process(a). The destination ToR $t(2,2)$ generates a meesage-response packet and send it to source ToR $t(1,2)$ who decides on the final flow to path allocation(b). }
\label{fig:E2}
\end{figure*}
\subsubsection{Failure Table (FT)}
The importance of failure table shines in asymmetric topology due to link failure. FTs are maintained only at aggregation switches and record their outport link failure. We leverage the structural property of 3-tier Clos topology to keep track of link failures that connect aggregate switches to the switches in the upstream tier with minimal overhead. In 3-tier Clos, aggregation switches that have the same ID in different pods are connected to each other through one core plane. Therefore, each entry record in a failure table maintain the status of the outport links belong to an aggregate switch in a  different pod. The table size is bounded  by the number of pods in the topology and the number of paths between the aggregation and the core layers. In 128-pod fat-tree with more than 524k hosts, there are 64 paths between aggregation and core switches. The status of each link is represented by a bit flag (1 for failed link, 0 for active link) where the index of a bit corresponds to the index of a port; for instance, the first bit corresponds to the first port in the switch and so on. Thus, 64 bits are enough to keep track of link failures in case of 128-pod fat-tree topology.

\subsubsection{Ethernet Tags}
In CAFT, congestion and link failure information are exchanged among switches during path allocation process using Ethernet tags on IP packets. We follow the tag format used in \cite{expeditus} that includes 4 fields as listed bellow:
\begin{enumerate}
  \item Tag protocol identifier (16 bits): This field identifies the packet as a path allocation process packet.
  \item Trip flag, TF (1 bit): This field indicates which way in the round-trip the packet serves (outbound(0) or inbound(1)). The tagged outbound packets are called "Message-request", and those for the inbound packets are called "Message-response".
  \item Number of congestion metrics, NoC (7 bits): This field indicates the number of carried congestion data bits in the packet.
  \item Congestion metrics, CM: This field holds the congestion data each represented in 4 bits with the first bit as a failure flag if set, otherwise it is ignored by switches.
\end{enumerate}

\subsection{Path Allocation Algorithm}
In this section we explain in details the allocation process of flows to paths in CAFT and show the efficiency of this mechanism in both symmetric and asymmetric toplogies. 
\subsubsection{Symmetry}
CAFT collects the complete congestion information of two candidate paths before making its final load balancing decision. The first path is chosen after exploring all possible paths through one core group during the SYN route from source to destination hosts. The second path, on the other hand, is selected out of all possible paths explored through different core group from destination to source hosts. Therefore, routing flows based on global congestion information minimizes their final completion time and shows CAFT's load balancing efficiency. A path in 3-tier Clos consists of four hops; therefore, the source ToR receives 8 congestion metrics at the end of the round-trip process. Figures~\ref{fig:E1} and~\ref{fig:E2} illustrate in details the processing steps of initiating two candidate paths and making the final path selection decision. There exists a new TCP connection between a host under ToR $t(1,2)$ and another host under ToR $t(2,2)$.
\begin{enumerate}
  \item Source ToR $t(1,2)$: Upon the arrival of the first packet of flow A (SYN), ToR tags the packet with its egress link loads of the uplinks (1,3) and sets TF to zero and NoC correspondingly. ToR also forwards the tagged packet, i.e. message-request, through the least congested link to the source aggregation switch $a(1,1)$ and creates a new table entry for flow A with outport set to the index of the least loaded link (1 in this example), valid bit set to 1 and the age bit set to 0.

  \item Source aggregation $a(1,1)$: Upon receiving the message-request packet, it adds its egress link congestion of the uplinks (4,2) to the received message-request packet. It forwards the message-request using ECMP and inserts a new entry to PAT with valid and age bits set to zero.

  \item Destination aggregation $a(2,1)$:  Upon receiving the message-request packet, it pulls the congestion information corresponding to the source aggregation $a(1,1)$, and aggregates it link by link with its ingress link loads (3,2) then it chooses the core switch with the minimum worst congestion. As shown in figure ~\ref{fig:E1}.a, it records the maximum of the two links connected to $c(1,1)$ which is 4 and the maximum of the two links connected to core $c(1,2)$ which is 2. Then it chooses the core switch that has min-max congestion which is $c(1,2)$ in this example.
       Two forwarding actions are triggered in this switch as shown in figure ~\ref{fig:E1}.b:
       First, it copies the TCP/IP header from the message-request and interchanges the source and destination IP addresses to generate a message-response with empty payload. It tags the message-response with TF set to zero and forwards it back to the source aggregation switch through the chosen core. The source aggregation reacts to the message response with an empty payload and a 0 TF value by interchanging the source/destination IP addresses, computing the flow ID, and recording the outport number for the flow.

       Second, it adds the congestion metrics of the links connected through the chosen core (2,2) to the message-request alongside the congestion information of the source ToR (1,2) and forwards it down to the destination ToR.

  \item Destination ToR $t(2,2)$:  Upon receiving the message-request packet, it pulls the congestion information corresponding to the source ToR (1,3), and aggregates it link by link with its ingress link loads to choose the aggregation switch with least min-worst link congestion $a(2,2)$ in this example. If the chosen aggregation happened to be the one who forwarded the message-request $a(2,1)$ in this example, then ToR chooses the switch with second min-worst link load. Therefore, the destination ToR switch will perform the following actions as shown in figure~\ref{fig:E2}.a: First, it generates a message-response packet to the source ToR $t(1,2)$, forwards it through the chosen switch $a(2,2)$, tagged with TF set to 1 and the congestion metrics of the chosen aggregation switch (3,0), the aggregation switch who forwarded the message-request to the destination ToR (1,4), and the chosen core switch (2,2) pulled from the received message-request. Second, it removes the message-request tag and forwards the packet to the destination host.
\end{enumerate}

The outbound journey of the message-request ended at the destination ToR completing the selection of the first candidate path and part of the second candidate path. The inbound journey involves a message-response packet traversing the network, carrying the selected paths' congestion information during the outbound trip, to complete the selection of the second candidate path. The response-message lands at the destination aggregation $a(2,2)$, as shown in figure ~\ref{fig:E2}.a, coming from destination ToR to explore the possible paths to the core tier. The aggregation tags the packet with its ingress link load metrics (3,3) and forwards it using ECMP. Upon receiving the message-response, source aggregation $a(1,2)$ pulls the ingress loads from the packet and aggregates them with its egress link loads (5,4) and chooses the core switch with min-worst congestion which is $c(2,2)$ in this example. Furthermore, it computes the flow's ID and inserts a new entry to PAT with outport set to best core port number. Moreover, it tags the message-response with the congestion metrics of the chosen core (3,4) and forwards it to the source ToR to make path selection decision. The flow is then assigned, by the source ToR, to the min-worst path, out of the two candidate paths, which is the blue path P1 as shown in figure~\ref{fig:E2}.b.
The packet processing logic of both ToR and aggregation switches is presented in algorithms 1 and 2.

\begin{figure}
	\centering{
\includegraphics[width=1\columnwidth]{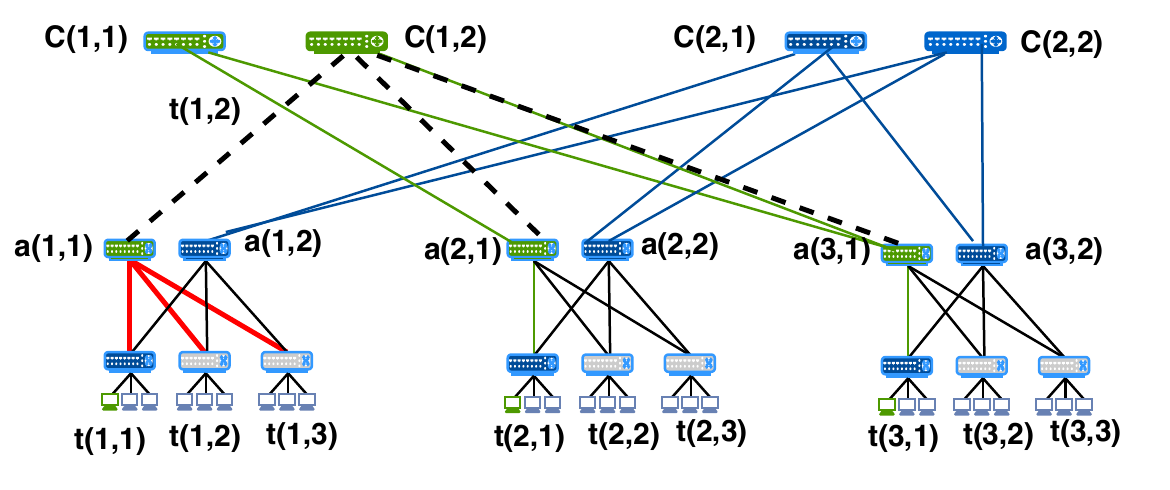}
	\caption{The link connects $a(1,1)$ and $c(1,1)$ fails. That causes all the links connected to $c(1,2)$ to be congested as they are the only routes to switch $a(1,1)$ in the first pod.}
\label{fig:failure}}
\end{figure}
\subsubsection{Asymmetry}
Asymmetries are the norm in large scale datacenter networks. Topology asymmetry can occur in many scenarios in datacenter such as when adding a new servers and switches of different types (e.g 10G and 40G core switches). Furthermore, it is reported that on average small datacenters experience more than 300 link failures each month~\cite{wcmp}. Moreover, large datacenters encounter frequent link cuts that create asymmetry in the topology \cite{cut1,cut2}. Handling asymmetry requires global knowledge about the downstream link status. Switches cannot balance flows by having only local link congestion information as it may cause more congestion due to the inefficient interaction with TCP's control loop. Next, using an example we show the efficiency of the proposed scheme, CAFT, in making load balancing decisions in case of topological asymmetry caused by link failures when compared to both congestion-agnostic (ECMP) and partial congestion-aware (Expeditus~\cite{expeditus}) schemes. Upon receiving a packet, ECMP routing protocol calculates a hash value based on the packet header fields that identify a flow and selects one of the equal cost path based on the hash value. On the other hand, Expeditus uses two stage path selection (look at figure~\ref{fig:failure}). In the first stage, the source ToR $t(2,1)$ congestion metrics is sent to the destination ToR $t(3,2)$ who selects the best aggregation switch (say $a(3,2)$). In the second stage, the chosen destination aggregation congestion metric is sent to source aggregation $a(2,2)$ to choose best core switch. The congestion information exchange is done via Exp-request and Exp-response packets. 

Consider the asymmetric scenario depicted in figure~\ref{fig:failure}. Suppose the link from $a(1,1)$ is failed, as a result, all the links connected to switch $c(1,2)$ are congested as they are the only routes to switch $a(1,1)$ in the first pod. Furthermore, all the uplinks from all ToR switches (red links) in the first pod to switch $a(1,1)$ will experience low load. Suppose there are flows from hosts in pod 2 to other hosts in pod 1, the $(a(2,1), c(1,2)$ link experience high congestion compared to the uplinks of $a(2,2)$ due to the link failure. Therefore, CAFT and Expeditus will route the traffic from switch $a(2,1)$ to other pods through core $(c(1,1)$ to avoid the congested link while ECMP distribute the traffic evenly and further congest $(a(2,1), c(1,2))$ link.

Yet, Expeditus is limited in case of assigning flows from pod 1 to other pods or vice versa due to the partial path knowledge of their two stage path selection algorithm.  Lets say there is a new flow from $t(1,1)$ to $t(2,1)$, Expeditus chooses best aggregation switch in stage 1 and best core switch in stage two. If $a(2,1)$ is chosen in stage 1, then there are two scenarios for chosen the best core in the second stage:
First if Exp-response packet routed through $c(1,1)$, then Exp-response packet will be dropped and one load balancing opportunity is lost. Second if the packet is routed through $c(1,2)$, then the flow will be assigned to a congested path. On the other hand, CAFT as explained in the previous section explores two candidate paths. In the first path source ToR $t(1,1)$ forwards the packet through least congested link(probably to $a(1,1)$ as it experience low load as discussed above). If $a(1,1)$ is selected then it adds the link failure information to the packet before forwarding it. Upon receiving the packet, $a(2,1)$ records the failing link information in a failure table (to avoid forwarding packets to pod 1 through $c(1,1)$ during link failure) then forwards the packet to destination ToR $t(2,1)$. The second packet will be routed through a blue path to reach $t(1,1)$ for final path selection decision. CAFT avoids the congested path by assigning the flow to the blue path as it has lower congestion compared to green path.

To further show the efficiency of the proposed load balancing protocol, consider another scenario where a flow is originated from pod 2 through $t(2,3)$ to $t(1,3)$ in pod 1. Expeditus uses ECMP to route the packet from $t(2,3)$ to one of the aggregation in the upper level. If the packet is forwarded to $a(2,1)$, it also uses ECMP to forward the packet to one of the core switches. If the packet is forwarded to $c(1,1)$ then the packet will be dropped and the load balancing opportunity will be lost. In the contrary, aggregation switch $a(2,1)$ using CAFT avoids forwarding the packet to core $c(1,1)$ by checking the failure table and finding the link failure information.

\section{\sc {Performance Evaluation}}
\label{sec:Evaluation}
In this section, we evaluate the performance of CAFT in large-scale data center networks by conducting several ns-3 simulations. The evaluation is based on realistic workload, derived from a large enterprise production data center that runs web search tasks~\cite{dctcp}. We generate traffic between senders and receivers in different pods randomly according to Poisson processes with varying arrival rates. We conduct our evaluation on 8-pod fat-tree \cite{fat} as the baseline topology. It contains 128 hosts that reside under 32 ToR switches. Each ToR is connected to 4 hosts at the downlink ports and 4 aggregation switches at the uplink ports. There are 16 core switches that connect aggregation switches in different pods, i.e. 16 equal cost paths between any two hosts in different pods. We set the links to run at 1Gbps and the fabric round trip time is set to $\SI{20}{\micro\second}$ across pods. 
TCP New Reno is used for the congestion control algorithm. The evaluation results are based on the average over 5 simulation runs each generating around 9K flows.

We compare the performance of CAFT with the following three schemes:
\begin{itemize}
  \item \textbf{Expeditus~\cite{expeditus}:} Distributed load balancing protocol for data centers that addresses the scalability challenges in general 3-tier Clos topologies that make a simple per-path congestion feedback approach impractical. Expeditus addresses these issues by collecting local congestion information, and applying  a two-stage path selection mechanism to carry congestion information across switches to make a flow routing decision.
  \item \textbf{Optimal:} It routes flows based on a complete end-to-end congestion information of all possible paths between any two hosts. This is used for baseline comparison. This is not practical to implement in real datacenters as it imposes significant overhead in 3-tier Clos topologies.
  \item \textbf{ECMP:} This is the commonly used routing in datacenters that provides equal-cost multipath.
\end{itemize}
\begin{figure}[h]
	\centering{
	\includegraphics[width=1\columnwidth,height=0.48\columnwidth]{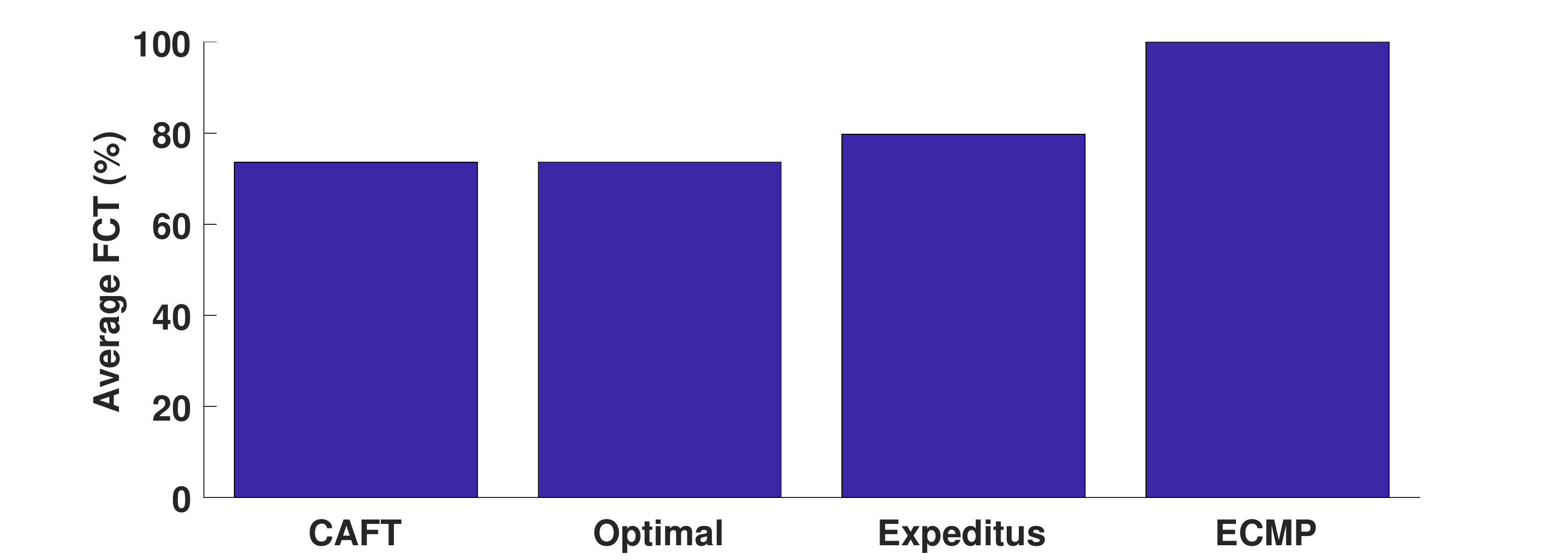}
	\caption{Average FCT for web search workload in the baseline 8-pod fat-tree with core tier oversubscription of 2:1}
	\label{fig:allFCT}}
\end{figure}
\subsection{Symmetric Topology}
\begin{figure}
	\centering{
	\includegraphics[width=1\columnwidth,height=0.48\columnwidth]{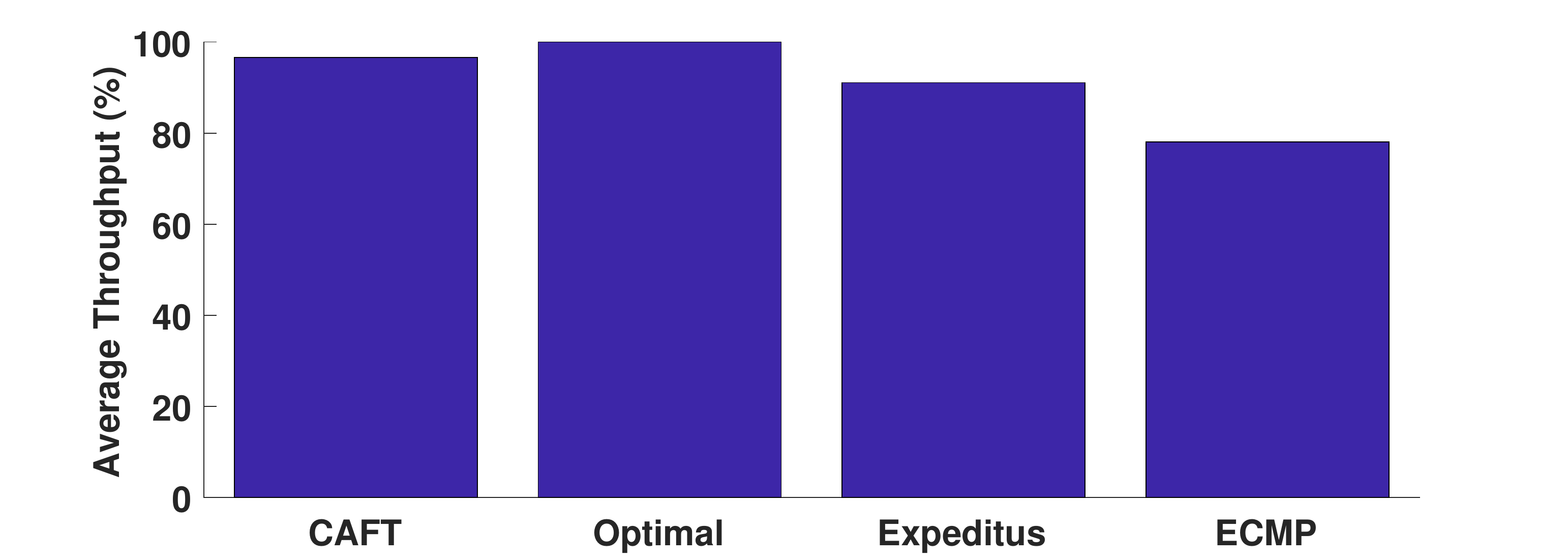}
	\caption{Average throughput for web search workload in the baseline 8-pod fat-tree with core tier oversubscription of 2:1}
	\label{fig:allPut}}
\end{figure}

 First, we evaluate the proposed scheme against existing schemes in terms of average throughput and flow final completion time (CFT). Figure \ref{fig:allFCT} shows the average flow completion time for all flows normalized to ECMP. Observe that CAFT has around 26$\%$ and 8$\%$ reduction compared to ECMP and Expeditus schemes while it achieves the same performance as the optimal. Figure \ref{fig:allPut} shows CAFT's throughput is very close to the optimal. ECMP suffers from low throughput when large flows are hashed to the same path.

\begin{figure}[h]
	\centering{
	\includegraphics[width=1\columnwidth,height=0.48\columnwidth]{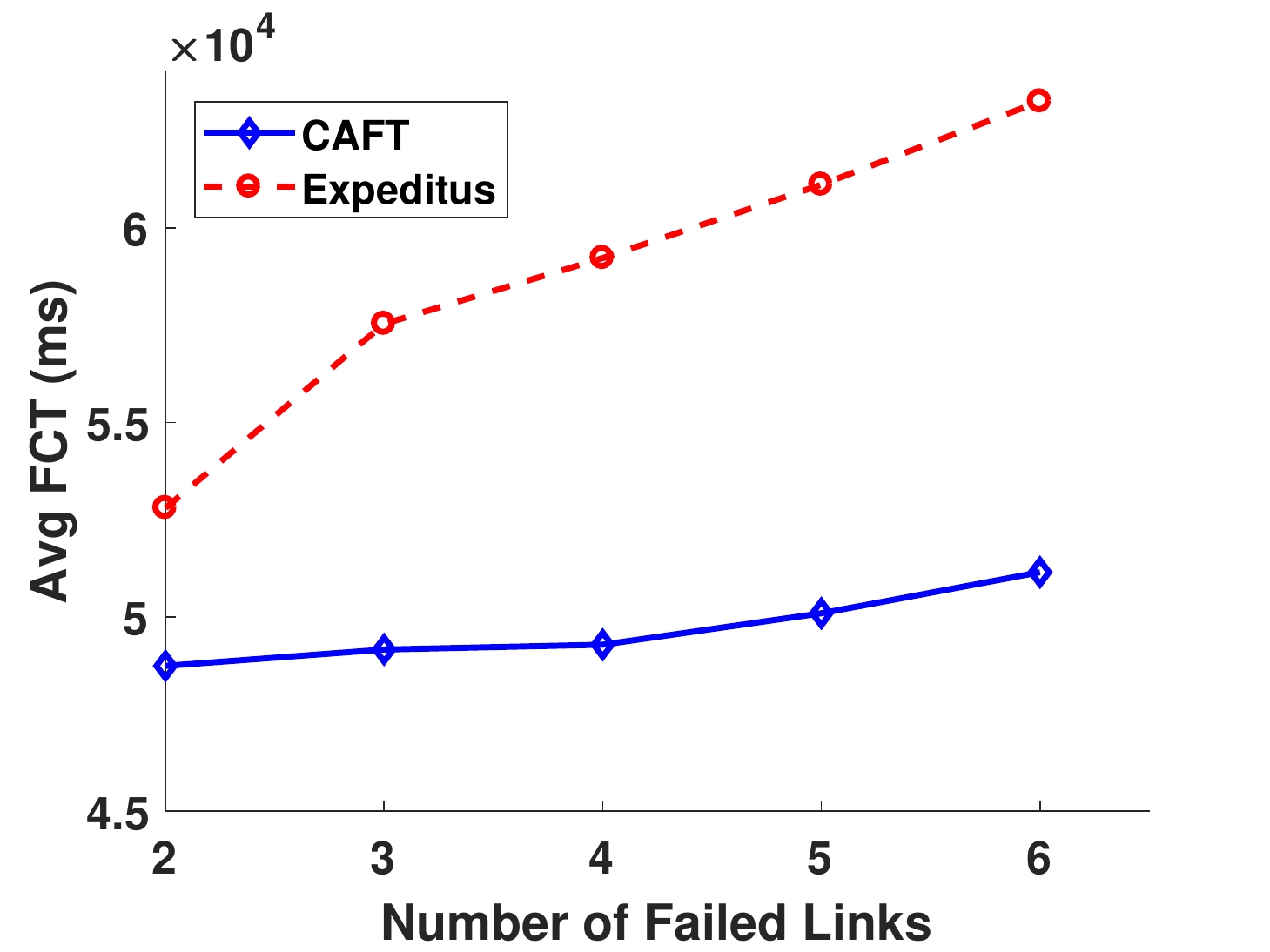}
	\caption{Average FCT for web search workload in the baseline 8-pod fat-tree with failures and core tier oversubscription of 2:1}
	\label{fig:f1}}
\end{figure}

\subsection{Topology Asymmetry}
In this section, we evaluate the impact of link failures and topology asymmetry on CAFT. We vary the number of failed links and choose links to fail uniformly at random. Figure \ref{fig:f1} shows the average FCT of both CAFT and Expeditus with failed links varying from $2$ to $6$. Observe that the performance gain of CAFT becomes more significant with more link failures reaching up to 20$\%$ reduction on flow FCT. This shows the efficiency of CAFT in minimizing flow FCT without imposing extra overhead. CAFT achieves this by maintaining the congestion information of two paths before making routing decisions.

%



\section{\sc {Conclusion}}
\label{sec:Conclusion}
We introduced CAFT, a congestion-aware fault-tolerant load balancing protocol for datacenters. CAFT explores two candidate paths during the TCP connection process and routes flows to the path with the minimum bottleneck link. We have evaluated CAFT through extensive ns-3 simulations, and demonstrated that CAFT is robust to asymmetry caused by link failures and outperforms existing schemes in terms of flow completion times and throughput. For future work, we plan to evaluate CAFT using data-mining workloads under different link failure scenarios. 
%

\bibliographystyle{IEEE}
\bibliography{References}

\begin{thebibliography}{10}

\bibitem{vl2}
Albert Greenberg, James~R Hamilton, Navendu Jain, Srikanth Kandula, Changhoon
  Kim, Parantap Lahiri, David~A Maltz, Parveen Patel, and Sudipta Sengupta,
\newblock ``Vl2: a scalable and flexible data center network,''
\newblock in {\em Proceedings of the ACM SIGCOMM 2009 conference on Data
  communication}, 2009, pp. 51--62.

\bibitem{fat}
Mohammad Al-Fares, Alexander Loukissas, and Amin Vahdat,
\newblock ``A scalable, commodity data center network architecture,''
\newblock {\em ACM SIGCOMM computer communication review}, vol. 38, no. 4, pp.
  63--74, 2008.

\bibitem{ecmp}
C~Hopps,
\newblock ``Rfc2992: analysis of an equal-cost multi-path algorithm,'' 2000.

\bibitem{hedera}
Mohammad Al-Fares, Sivasankar Radhakrishnan, Barath Raghavan, Nelson Huang,
  Amin Vahdat, et~al.,
\newblock ``Hedera: dynamic flow scheduling for data center networks.,''
\newblock in {\em Nsdi}, 2010, vol.~10, pp. 89--92.

\bibitem{mptcp}
Damon Wischik, Costin Raiciu, Adam Greenhalgh, and Mark Handley,
\newblock ``Design, implementation and evaluation of congestion control for
  multipath tcp.,''
\newblock in {\em NSDI}, 2011, vol.~11, pp. 8--8.

\bibitem{conga}
Mohammad Alizadeh, Tom Edsall, Sarang Dharmapurikar, Ramanan Vaidyanathan,
  Kevin Chu, Andy Fingerhut, Francis Matus, Rong Pan, Navindra Yadav, George
  Varghese, et~al.,
\newblock ``Conga: Distributed congestion-aware load balancing for
  datacenters,''
\newblock in {\em ACM SIGCOMM Computer Communication Review}. ACM, 2014, pp.
  503--514.

\bibitem{expeditus}
Peng Wang, Hong Xu, Zhixiong Niu, Dongsu Han, and Yongqiang Xiong,
\newblock ``Expeditus: Congestion-aware load balancing in clos data center
  networks,''
\newblock in {\em Proceedings of the Seventh ACM Symposium on Cloud Computing}.
  ACM, 2016, pp. 442--455.

\bibitem{wcmp}
Junlan Zhou, Malveeka Tewari, Min Zhu, Abdul Kabbani, Leon Poutievski, Arjun
  Singh, and Amin Vahdat,
\newblock ``Wcmp: Weighted cost multipathing for improved fairness in data
  centers,''
\newblock in {\em Proceedings of the Ninth European Conference on Computer
  Systems}. ACM, 2014, p.~5.

\bibitem{cut1}
Phillipa Gill, Navendu Jain, and Nachiappan Nagappan,
\newblock ``Understanding network failures in data centers: measurement,
  analysis, and implications,''
\newblock {\em ACM SIGCOMM Computer Communication Review}, vol. 41, no. 4, pp.
  350--361, 2011.

\bibitem{cut2}
Chuanxiong Guo, Lihua Yuan, Dong Xiang, Yingnong Dang, Ray Huang, Dave Maltz,
  Zhaoyi Liu, Vin Wang, Bin Pang, Hua Chen, et~al.,
\newblock ``Pingmesh: A large-scale system for data center network latency
  measurement and analysis,''
\newblock in {\em ACM SIGCOMM Computer Communication Review}. ACM, 2015, pp.
  139--152.

\bibitem{dctcp}
Mohammad Alizadeh, Albert Greenberg, David~A Maltz, Jitendra Padhye, Parveen
  Patel, Balaji Prabhakar, Sudipta Sengupta, and Murari Sridharan,
\newblock ``Data center tcp (dctcp),''
\newblock in {\em Proceedings of the ACM SIGCOMM 2010 conference}, 2010, pp.
  63--74.

\end{thebibliography}
\end{document}